\documentclass[prb,twocolumn]{revtex4-1} 

\usepackage{amsmath}  
\usepackage{amsfonts} 
\usepackage{graphicx} 
\usepackage{amssymb}
\usepackage[utf8]{inputenc}
\usepackage{dcolumn}
\usepackage{bm}
\usepackage{graphicx}

\newcommand{\diff}{\text{d}}

\newcommand{\vek}[1]{\mathbf{#1}}
\newcommand{\vekh}[1]{\mathbf{\hat{#1}}}

\renewcommand{\div}[0]{\nabla\cdot\vek}

\newcommand{\curl}{\nabla\times\vek}

\newcommand{\be}{\begin{equation}}
\newcommand{\ee}{\end{equation}}

\begin{document}


\title{Dielectric media considered as vacuum with sources}

\author{Hans Olaf Hågenvik}
\affiliation{Department of Electronic Systems, NTNU -- Norwegian University of Science and Technology, NO-7491 Trondheim, Norway}

\author{Kjell Bløtekjær}
\affiliation{Department of Electronic Systems, NTNU -- Norwegian University of Science and Technology, NO-7491 Trondheim, Norway}

\author{Johannes Skaar}
\affiliation{Department of Electronic Systems, NTNU -- Norwegian University of Science and Technology, NO-7491 Trondheim, Norway}
\affiliation{Department of Technology Systems, University of Oslo, Box 70, NO-2027 Kjeller, Norway}


\date{\today}

\begin{abstract}
Conventional textbook treatments on electromagnetic wave propagation consider the induced charge and current densities as ``bound'', and therefore absorb them into a refractive index. In principle it must also be possible to treat the medium as vacuum, but with explicit charge and current densities. This gives a more direct, physical description. However, since the induced waves propagate in vacuum in this picture, it is not straightforward to realize that the wavelength becomes different compared to that in vacuum. We provide an explanation, and also associated time-domain simulations. As an extra bonus the results turn out to illuminate the behavior of metamaterials.
\end{abstract}

\maketitle 

\section{Introduction}

Electromagnetic fields in a medium are governed by Maxwell's equations
\begin{subequations}\label{ME}
\begin{align}
	\nabla\cdot \vek D &= \rho_\text{free}, \\
	\nabla\cdot \vek B &=0, \label{divB0}\\
	\curl{E} &= i\omega\vek B, \label{Faraday}\\
	\curl H &= -i\omega \vek D + \vek J_\text{free}.
\end{align}
\end{subequations}
Here we have assumed harmonic time dependence $\exp(-i\omega t)$. The two equations containing the auxiliary fields $\vek D$ and $\vek H$ are however not unique. Their forms are dependent on which charges are consider as ``bound'' and which are ``free''. Taking a source-free dielectric medium as an example, the charge and current densities are conventionally considered as ``bound'', and therefore described by the polarization density $\vek P$. In this way, the charges and currents do not bother us; they are simply taken into account by using a (possibly complex) permittivity $\epsilon$. In this picture, the two equations read
\begin{subequations}\label{dielpict}
\begin{align}
\nabla\cdot \epsilon\vek E &= 0 , \\
\mu_0^{-1}\curl B &= -i\omega\epsilon\vek E.
\end{align}
\end{subequations}

There is another possibility: The medium can be considered as vacuum, with charge and current densities described explicitly. In this picture, we have
\begin{subequations}\label{vacpict}
\begin{align}
\nabla\cdot \epsilon_0\vek E &= \rho , \\
\mu_0^{-1}\curl B &= -i\omega\epsilon_0\vek E + \vek J,
\end{align}
\end{subequations}
rather than \eqref{dielpict}. If we write $\epsilon\vek E=\epsilon_0\vek E+\vek P$ in \eqref{dielpict}, we find the connection between the two pictures:
\begin{subequations}\label{connpict}
\begin{align}
\vek J &=-i\omega\vek P, \\
\rho &= -\div P.
\end{align}
\end{subequations}

In this note, we would like to use the two pictures to solve Maxwell's equations in a dielectric slab. As we know from basic physics, the phase velocity and wavelength in the medium will be different from the values in vacuum \cite{landau_lifshitz_edcm,jackson,griffiths}. In the first picture, this is shown straightforwardly in terms of a refractive index different from unity. In the second picture, however, it is less straightforward to realize that the phase velocity and wavelength are changed compared to the situation in vacuum. After all, in this picture the medium is considered as vacuum, with additional current and charge density which can be viewed as sources. One would think that a superposition of waves with vacuum wavelength will result in the same vacuum wavelength.

It is worth mentioning that the second picture does not describe any new physical effects compared to the first picture. While the mathematical reason for the altered wavelength is clear in the first picture (Sec. \ref{sec:bound}), the physical reason will become more transparent in the second (Sec. \ref{sec:free}). The fact that the induced field exactly cancels the incident field is known as the Ewald--Oseen extinction theorem \cite{bornwolf,fearn96,ballenegger99}.

When considering wave propagation in dielectric media, it is usually most practical to consider the induced charges and currents as bound. The simplicity of the explanation of the altered wavelength and phase velocity in terms of a different refractive index is definitely appealing, compared to the more complicated, but perhaps more illuminating, explanation in the second picture. However, in some circumstances it is natural to consider induced currents as free, e.g., the induced current in a coil of wire due to a time varying magnetic field. 

Whether induced currents should be characterized as free or bound is a relevant question within the research field of metamaterials. By designing structures with characteristic size small compared to the wavelength of the electromagnetic radiation, one may control the induced currents to obtain electromagnetic responses not available in natural materials. The superposition of fields produced by the induced currents in the designed structures gives effective permittivity $\epsilon_\text{eff}$ and permeability $\mu_\text{eff}$. 

As an example, consider a metamaterial consisting of periodically arranged, parallel, thin metallic wires surrounded by vacuum \cite{pendry98}. For large wavelengths compared to the lattice constant, the metamaterial can be viewed as a continuous plasma with negative effective permittivity. This corresponds to picture 1, since the currents in the wires are absorbed into an effective permittivity. For sufficiently small wavelengths it makes little sense to describe the structure using an effective permittivity; it is more practical to describe the currents in the wires explicitly (picture 2). In an intermediate range of wavelengths, the medium can be described with an effective permittivity, however with spatial dispersion. In this range both pictures can be useful, dependent on the particular application.

We will consider a simple metamaterial example: 1d propagation through a periodic, layered structure \cite{saleh} of alternating permittivities $\epsilon_1$ and $\epsilon_2$. The effective permittivity becomes a weighted average of the two permittivities:
\begin{equation}\label{epseff}
   \epsilon_\text{eff} = \frac{\epsilon_1d_1 + \epsilon_2 d_2}{d_1 + d_2}.
\end{equation}
Here $d_1$ and $d_2$ are the thicknesses of the two alternating layers. Such a metamaterial structure is used to visualize the second picture through simulations in Sec. \ref{sec:fdtd}. 

This paper is intended for teachers and students in undergraduate physics, familiar with basic electromagnetic fields and waves. In particular, the discussion illuminates the freedom in Maxwell's equations when it comes to whether charges and currents are considered as ``free`` or ``bound``. This provides a connection between undergraduate physics education and modern metamaterial research.

\section{Bound charges and currents}\label{sec:bound}
\begin{figure}
\begin{center}
\includegraphics[width=\columnwidth]{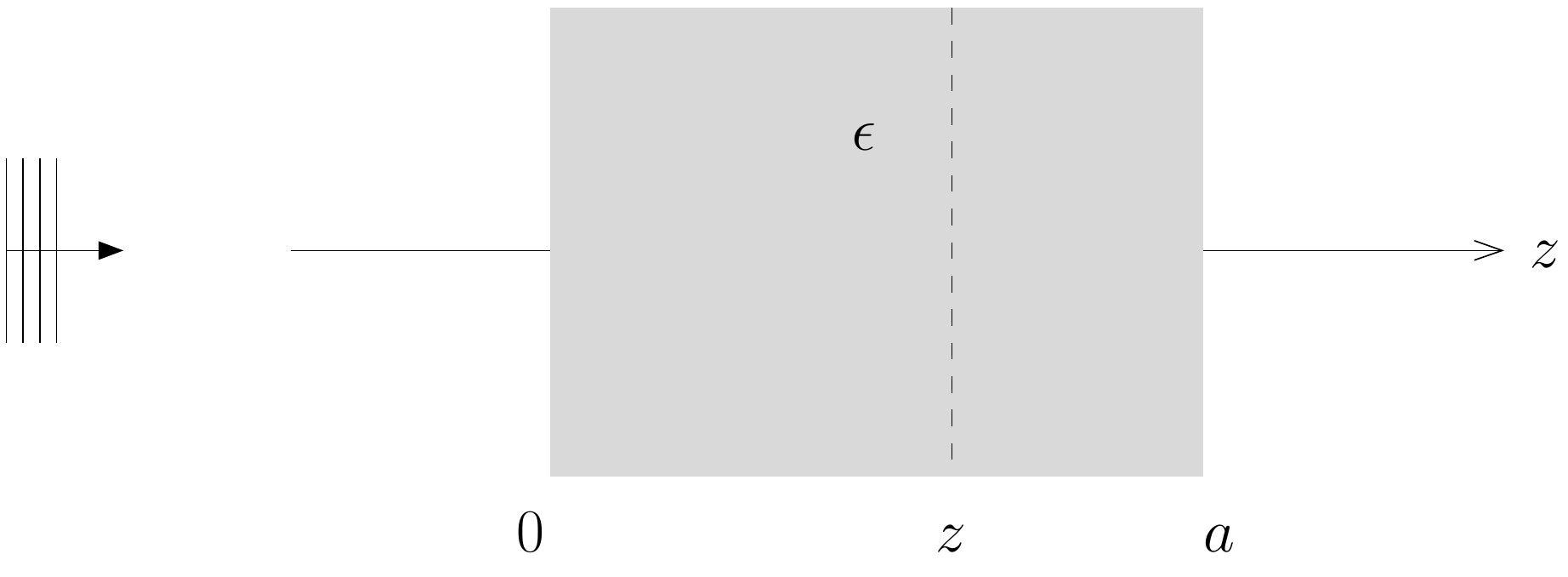}
\caption{A dielectric slab located in the region $0<z<a$. A plane wave is incident from $z=-\infty$.}
\label{fig:slab}
\end{center}
\end{figure}

We consider a dielectric slab, located in the region $0<z<a$, with vacuum elsewhere, see Fig. \ref{fig:slab}. The dielectric medium is assumed to be linear, isotropic, and homogeneous. A plane wave is normally incident from a source located at $z=-\infty$. From Maxwell's equations we can derive Helmholtz' equation,
\be
E''(z)+\beta^2 E(z) = 0,
\ee
with the following field solution in the dielectric medium:
\be\label{Efelt}
E(z)=Ae^{i\beta z}+Be^{-i\beta z}.
\ee
Here $\beta=kn$, $n=\sqrt{\epsilon_\text{r}}$ is the refractive index, $\epsilon_\text{r}=\epsilon/\epsilon_0$ is the relative permittivity, $k=\omega/c$ is the vacuum wavenumber, and $c$ is the vacuum light velocity. From \eqref{Efelt} we realize that the wavelength in the medium is $2\pi/\beta=(2\pi/k)/n=\lambda/n$, and the phase velocity is $c/n$. Here $\lambda=2\pi/k$ is the vacuum wavelength. The constants $A$ and $B$ can be determined from $a$, $\epsilon$, and the source by using the electromagnetic boundary conditions; that the electric and magnetic fields must be continuous at $z=0$ and $z=a$. 

\section{Free charges and currents}\label{sec:free}
We would like to explain the result in the previous section by considering the medium as charges and currents situated in a vacuum. In each plane $z'$ in the material, there is a current density $J(z')$, while the charge density is zero, according to \eqref{connpict}. The current density can be viewed as a source, distributed over the volume of the slab. The current generates a varying magnetic field and therefore also an electric field. By solving Maxwell's equations for a current source plane with surface current density $J(z')\diff z'$ in vacuum, we obtain the field
\be
\diff E_J(z,z') =
\begin{cases}
&-\frac{\eta}{2} J(z') e^{ik(z-z')} \diff z' \text{ for $z>z'$}, \\ 
&-\frac{\eta}{2} J(z') e^{-ik(z-z')} \diff z' \text{ for $z<z'$},
\end{cases}
\ee
where $\eta=\sqrt{\mu_0/\epsilon_0}$ is the wave impedance in vacuum (see Appendix \ref{sec:currentsheet} or e.g. Ref. \cite{pozar}). In the observation plane $z$, the total electric field will be a superposition of the field from all current planes, in addition to the direct field $E_\text{s}(z)$ from the source. If we let $0<z<a$, we find
\begin{align}
E(z) = E_\text{s}(z) &- \frac{\eta}{2}\int_{0}^z J(z')e^{ik(z-z')}\diff z' \nonumber\\ 
&- \frac{\eta}{2}\int_{z}^a J(z')e^{ik(z'-z)}\diff z'.
\end{align}
To proceed, we need to know the connection between the electric field and the resulting current. Assuming $J=-i\omega\epsilon_0\chi E$ for a susceptibility $\chi$,
\begin{align}\label{inteq}
E(z)= E_\text{s}(z) &+ \xi e^{ikz}\int_{0}^z E(z')e^{-ikz'}\diff z' \nonumber\\ 
&+ \xi e^{-ikz}\int_{z}^a E(z')e^{ikz'}\diff z',
\end{align}
where $\xi=i\omega\epsilon_0\chi\eta/2=i k\chi/2$. This integral equation expresses the total electric field as a sum of the source field and the induced field from the currents in the medium. 

To solve the integral equation, we may differentiate \eqref{inteq} twice, and substitute the original equation to eliminate the integrals. Taking the source to be $E_\text{s}(z)=\exp(ikz)$, the result is
\be\label{diffpict2}
E''(z)+\beta^2 E(z) = 0,
\ee
where $\beta$ now is defined by 
\be
\beta^2=k^2\left(1-2i\xi/k\right)=k^2\left(1+\chi\right).
\ee
To get correspondence between the two pictures, we need that $\epsilon_\text{r}=1+\chi$, which is recognized as the usual connection between susceptibility and permittivity \cite{landau_lifshitz_edcm,jackson}.

The general solution to \eqref{diffpict2} is given by \eqref{Efelt}. The constants $A$ and $B$ cannot be determined from boundary conditions as in Sec. \ref{sec:bound}, since now there is vacuum everywhere. However, they can be determined by the integral equation \eqref{inteq} directly. Substitute \eqref{Efelt} back into \eqref{inteq}:
\begin{align}
E(z) = Ae^{i\beta z} &+ Be^{-i\beta z} \nonumber\\
= e^{ikz} &+ \xi A e^{ikz}\int_{0}^z e^{i\beta z'-ikz'}\diff z' \nonumber\\ 
&+ \xi B e^{ikz}\int_{0}^z e^{-i\beta z'-ikz'}\diff z' \nonumber\\
&+ \xi A e^{-ikz}\int_{z}^a e^{i\beta z'+ikz'}\diff z' \nonumber\\
&+ \xi B e^{-ikz}\int_{z}^a e^{-i\beta z'+ ikz'}\diff z'.
\end{align}
By calculating the integrals, one finds that the coefficients of the resulting $\exp(\pm i\beta z)$ terms balance on each side of the equation. Once the $\exp(\pm i\beta z)$ terms are removed, $A$ and $B$ are found so as to make the remaining $\exp(\pm ikz)$ terms exactly cancel. In particular, it then turns out that the $z$-dependence of the integral terms cancels the $\exp(ikz)$ dependence of the source.

The reason for the dependence $e^{\pm i\beta z}$ in the medium, rather than $e^{\pm ikz}$, can be explained intuitively as follows. First, let the observation plane $z$ be outside the slab, i.e., $z>a$. Rather than \eqref{inteq}, we would then have
\be
E(z)= E_\text{s}(z) + \xi e^{ikz}\int_{0}^a E(z')e^{-ikz'}\diff z',
\ee
since now, the observation plane is located to the right of all current sources ($z>z'$). Similarly, for $z<0$ we have
\be
E(z)= E_\text{s}(z) + \xi e^{-ikz}\int_{0}^a E(z')e^{ikz'}\diff z'.
\ee
Clearly, all sources (or induced currents) generate waves of the form $e^{\pm ikz}$, which after superposition also can be written in this form. 

Returning to an observation plane inside the slab, we must use \eqref{inteq}. As we move the observation plane to the right, a different set of sources will contribute to the forward-propagating wave, as seen by the upper limit $z$ in the first integral. This means that the $z$-dependence is not only a result of the $z$-dependence of each wave separately, but also a result of the fact that the set of contributing sources to the left of the observation plane is dependent on the position of the observation plane.

This provides a different perspective compared to the analysis by James and Griffiths \cite{griffiths}, who viewed the response of the dielectric medium as a perturbation expansion: The vacuum electric field induces a polarization, which induces a field, which in turn induces another polarization, etc. While such a perturbation series converges to a wave with reduced speed $c/n$, the analysis is somewhat complicated and does not explain the physical mechanism for the altered wavelength in terms of $z$-dependent sets of sources.

\section{FDTD simulations}\label{sec:fdtd}
\begin{figure}[t!]
	\centering
    \includegraphics[width=\columnwidth]{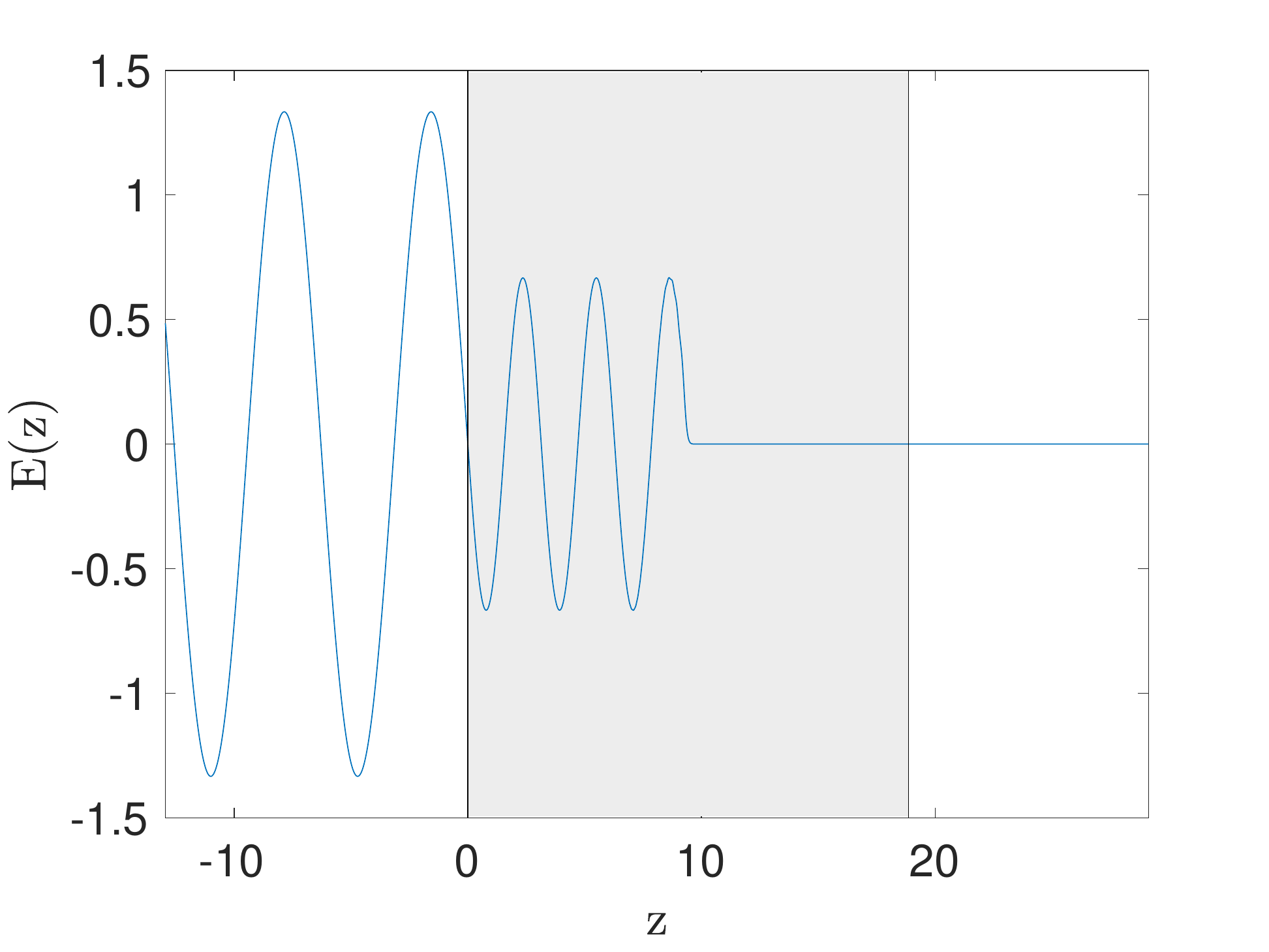}
    \caption{A plane wave is incident from a source located at $z=-\infty$, propagating through a dielectric slab with constant permittivity $\epsilon_\text{r}=4$. Supplementary video will accompany the article.}
    \label{fig:fdtd_a}
\end{figure}

\begin{figure}[t!]
	\centering
    \includegraphics[width=\columnwidth]{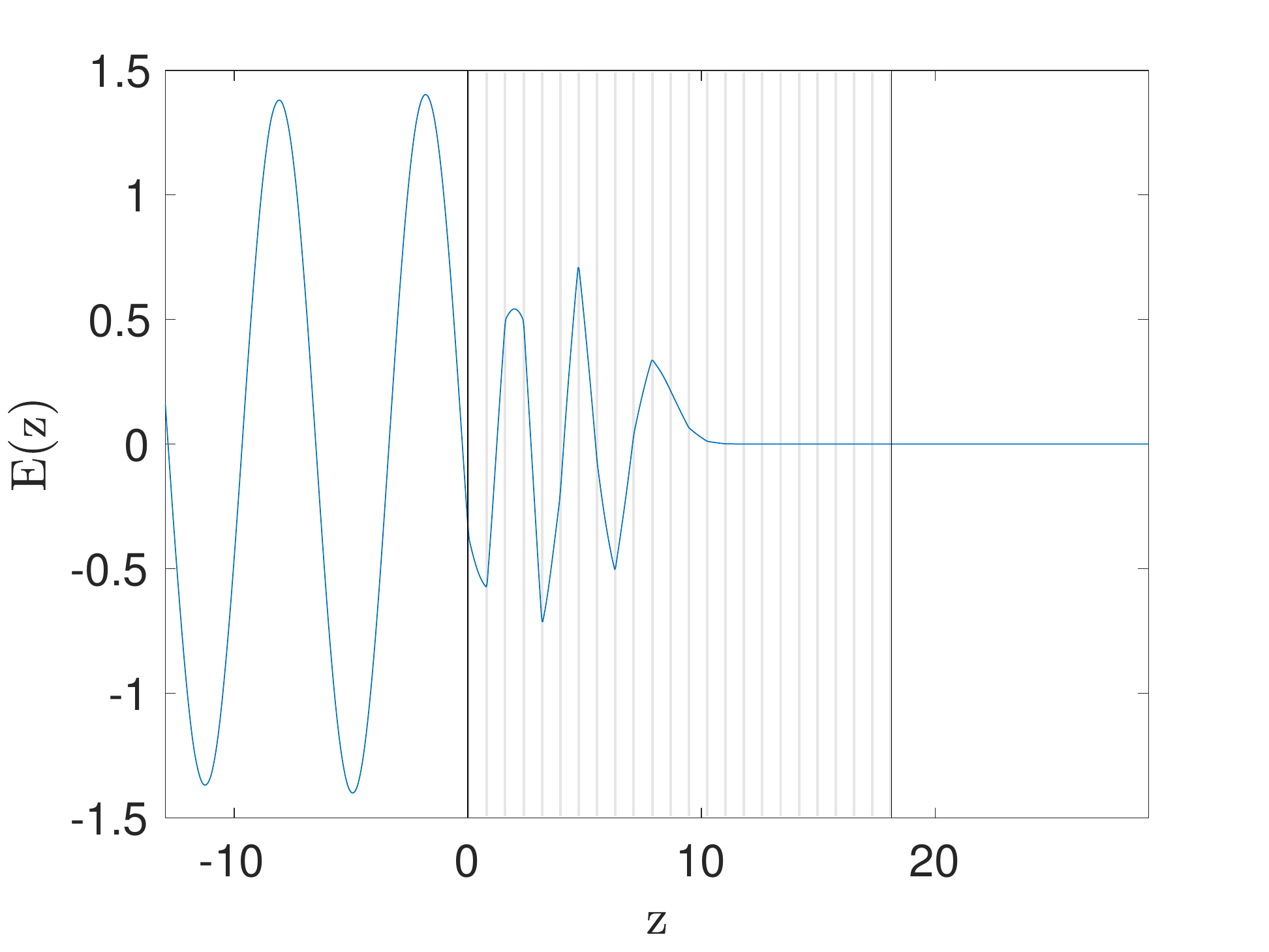}
    \caption{A plane wave is incident from a source located at $z=-\infty$, propagating through a metamaterial slab with $\epsilon_\text{eff}=4$. The slab is a layered structure where a high-index medium with $\epsilon_2 = 31$ fills $10$\% of the slab volume; the remaining layers consist of vacuum ($\epsilon_1 = 1$). Supplementary video will accompany the article.}
    \label{fig:fdtd_c}
\end{figure}

Figures \ref{fig:fdtd_a} and \ref{fig:fdtd_c} show the resulting electric field of an electromagnetic wave propagating through a slab. We have used Finite Difference Time Domain (FDTD) simulations \cite{yee66,kunz93,schneider10} of the conventional Maxwell equations in a dielectric medium, i.e., the time-domain counterparts of \eqref{divB0}, \eqref{Faraday}, and \eqref{dielpict}. Outside the slab there is vacuum. In both simulations normalized units have been used. The source produces a wave with frequency $\omega = 1$ approaching the slab from the left. The speed of light in vacuum is taken to be $c=1$. Using these units the vacuum wavelength at $\omega=1$ is $\lambda=2\pi$.

In Fig. \ref{fig:fdtd_a} the slab consists of a dielectric medium with $\epsilon_\text{r} = 4$. This gives a refractive index $n=\sqrt{\epsilon_\text{r}} = 2$. The wave propagates at the speed ${c}/{2}$, and the wavelength inside the slab is $\lambda/2$, in agreement with the first picture (Sec. \ref{sec:bound}). 


In Fig. \ref{fig:fdtd_c} the slab is a composite medium or metamaterial. The layered structure consists of a high-index medium ($\epsilon_2=31$) which fills $10$\% of the slab volume. Between the high-index layers there is vacuum ($\epsilon_1 = 1$). According to \eqref{epseff} the homogenized, effective permittivity is $\epsilon_{\text{eff}}=4$. The high-index layers are distributed evenly throughout the slab, with a separation distance such that there are approximately $4$ units cells over one effective wavelength. In other words, we have essentially compressed the induced currents in the homogeneous slab into thin current sheets surrounded by vacuum. 

In some sense Fig. \ref{fig:fdtd_c} therefore visualizes the second picture (Sec. \ref{sec:free}): The total electric field is a superposition of the source field, and the fields produced by induced currents in all high-index layers. When viewing the field in Fig. \ref{fig:fdtd_c} in detail, one finds that the variation with $z$ in the vacuum layers are slow, corresponding to waves in a vacuum, while the variation in the high-index layers are rapid. However, the resulting wave, if the small features are washed out, is approximately as in Fig. \ref{fig:fdtd_a}. In the present simulation, the parameters were chosen such that the nonideal effects are visible, to see the behavior in the different layers. A more homogeneous solution could be obtained by spreading the high-index layers out, i.e., more units cells per wavelength.

\section{Conclusion}
We have described propagation through a dielectric slab using two different pictures. The first picture, which is the conventional one, regards the induced charges and currents as bound, conveniently absorbing them into a relative permittivity $\epsilon_\text{r}$. The fact that the wavelength and phase velocity of the electromagnetic wave is different in the medium compared to the situation in vacuum is explained in terms of a refractive index $n=\sqrt{\epsilon_\text{r}}$.

In the second picture, the medium is instead considered as vacuum, with source charge and current densities. A superposition of waves from sources in vacuum seems to imply a wavelength equal to the vacuum wavelength. However, by examining the superposition in detail, we find that the altered wavelength and propagation speed is a result of the fact that the set of sources to the left and right of an observation plane $z$ depends on $z$. Although the calculations become more complicated, this picture provides useful physical insights.

Figures and animations of the resulting electric field from FDTD simulations are provided to visualize the two descriptions. The second picture together with the simulation of the metamaterial slab, may also be useful when it comes to understanding the homogenized, effective parameters of metamaterials.

\appendix
\section{Field from a uniform surface current source}\label{sec:currentsheet}
\begin{figure}[bh]
\begin{center}
\includegraphics[width=\columnwidth]{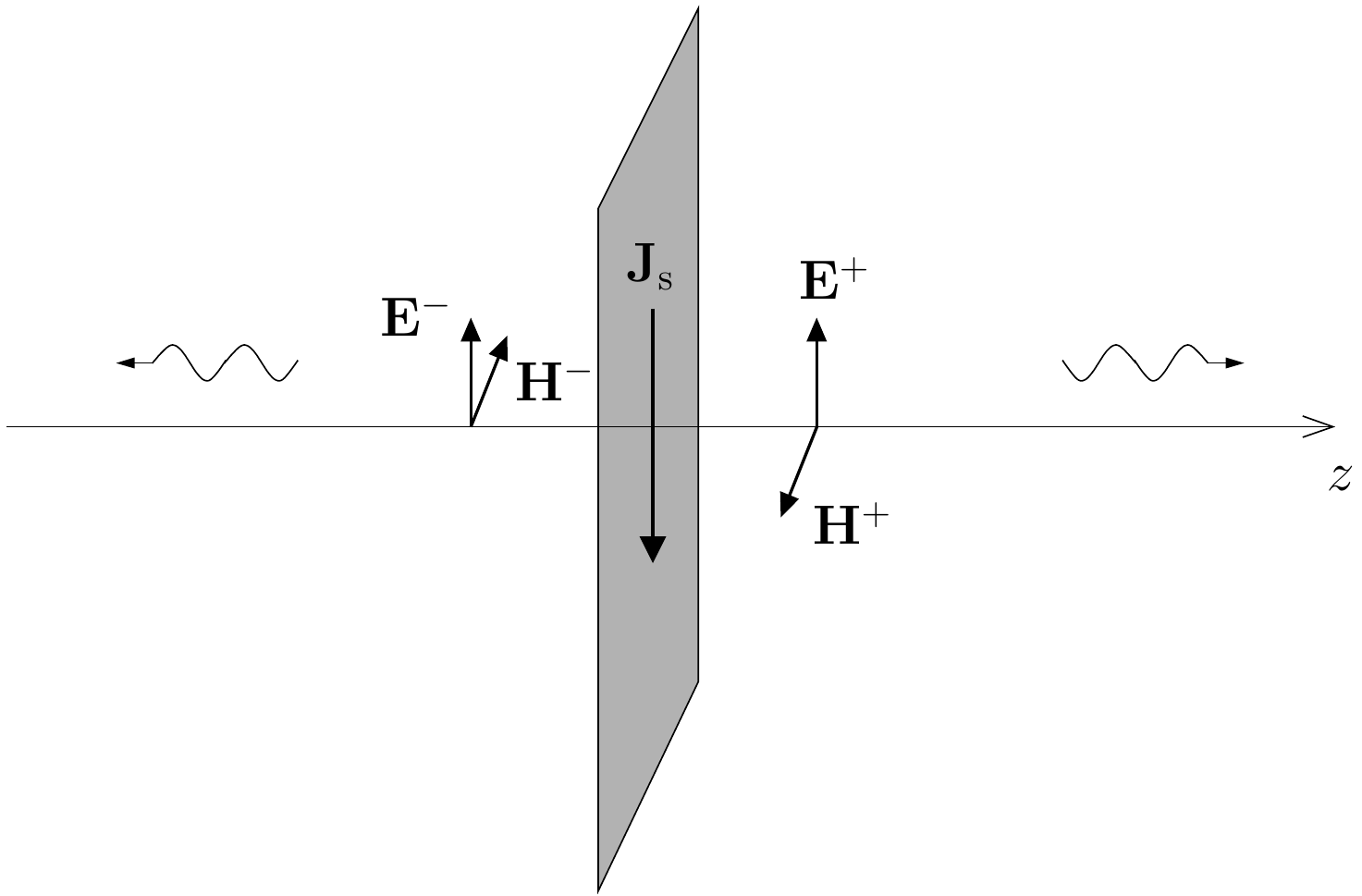}
\caption{A uniform surface current source $\vek J_\text{s}$ is located in the plane $z=0$. A harmonic time dependence $\exp(-i\omega t)$ is assumed.}
\label{fig:currentsource}
\end{center}
\end{figure}
Let a uniform surface current source $\vek J_\text{s}$ be located in the plane $z=0$, and surrounded by vacuum, see Fig. \ref{fig:currentsource}. We want to determine the electric field everywhere. In the region $z>0$, the electric field solution to Helmholtz' equation must be of the type $\vek E^+\exp(ikz)$, while in the region $z<0$ we must have a field $\vek E^-\exp(-ikz)$. From reflection symmetry about the plane $z=0$, $\vek E^-=\vek E^+$. We choose the coordinate system such that $\vek E^+=E^+\vekh x$. From Faraday's law this gives a magnetic field
\be
\vek H^+ = \frac{1}{i\omega\mu_0} \curl E^+ = \frac{kE^+}{\omega\mu_0}\vekh y = \frac{E^+}{\eta}\vekh y.
\ee
Similarly we find $\vek H^-=-\vek H^+$. Now we apply the Maxwell boundary condition 
\be
\vek H^+-\vek H^-=\vek J_\text{s}\times\vekh z,
\ee
to determine the constant $\vek E^+$. This gives $\vek E^+=-\frac{\eta}{2}\vek J_\text{s}$, and therefore
\be
\vek E(z) =
\begin{cases}
&-\frac{\eta}{2} \vek J_\text{s} e^{ikz} \text{ for $z>0$}, \\ 
&-\frac{\eta}{2} \vek J_\text{s} e^{-ikz} \text{ for $z<0$}.
\end{cases}
\ee

\def\cprime{$'$}

\end{document}